\documentclass[twocolumn]{aa}

\usepackage{graphicx}
\usepackage{amsmath,amsfonts,amssymb}
\usepackage{txfonts}
\usepackage{color}
\usepackage{natbib}
\usepackage{float}
\usepackage{dblfloatfix}
\usepackage{afterpage}
\usepackage{ifthen}
\usepackage[morefloats=12]{morefloats}
\usepackage{placeins}
\usepackage{multicol}
\bibpunct{(}{)}{;}{a}{}{,}
\usepackage[switch]{lineno}
\definecolor{linkcolor}{rgb}{0.6,0,0}
\definecolor{citecolor}{rgb}{0,0,0.75}
\definecolor{urlcolor}{rgb}{0.12,0.46,0.7}
\usepackage[breaklinks, colorlinks, urlcolor=urlcolor,
    linkcolor=linkcolor,citecolor=citecolor,pdfencoding=auto]{hyperref}
\hypersetup{linktocpage}

\def\setsymbol#1#2{\expandafter\def\csname #1\endcsname{#2}}
\def\getsymbol#1{\csname #1\endcsname}

\def\Planck{\textit{Planck}}

\newbox\tablebox    \newdimen\tablewidth
\def\leaderfil{\leaders\hbox to 5pt{\hss.\hss}\hfil}
\def\tablenote#1 #2\par{\begingroup \parindent=0.8em
    \abovedisplayshortskip=0pt\belowdisplayshortskip=0pt
    \noindent
    $$\hss\vbox{\hsize\tablewidth \hangindent=\parindent \hangafter=1 \noindent
    \hbox to \parindent{$^#1$\hss}\strut#2\strut\par}\hss$$
    \endgroup}

\def\L2{\ifmmode L_2\else $L_2$\fi}

\def\DeltaT{\ifmmode \Delta T\else $\Delta T$\fi}
\def\deltat{\ifmmode \Delta t\else $\Delta t$\fi}
\def\fknee{\ifmmode f_{\rm knee}\else $f_{\rm knee}$\fi}
\def\Fmax{\ifmmode F_{\rm max}\else $F_{\rm max}$\fi}
\def\solar{\ifmmode{\rm M}_{\mathord\odot}\else${\rm M}_{\mathord\odot}$\fi}
\def\Msolar{\ifmmode{\rm M}_{\mathord\odot}\else${\rm M}_{\mathord\odot}$\fi}
\def\Lsolar{\ifmmode{\rm L}_{\mathord\odot}\else${\rm L}_{\mathord\odot}$\fi}
\def\inv{\ifmmode^{-1}\else$^{-1}$\fi}
\def\mo{\ifmmode^{-1}\else$^{-1}$\fi}
\def\sup#1{\ifmmode ^{\rm #1}\else $^{\rm #1}$\fi}
\def\expo#1{\ifmmode \times 10^{#1}\else $\times 10^{#1}$\fi}
\def\,{\thinspace}
\def\lsim{\mathrel{\raise .4ex\hbox{\rlap{$<$}\lower 1.2ex\hbox{$\sim$}}}}
\def\gsim{\mathrel{\raise .4ex\hbox{\rlap{$>$}\lower 1.2ex\hbox{$\sim$}}}}

\def\simprop{\mathrel{\raise .4ex\hbox{\rlap{$\propto$}\lower 1.2ex\hbox{$\sim$}}}}
\def\deg{\ifmmode^\circ\else$^\circ$\fi}
\def\pdeg{\ifmmode $\setbox0=\hbox{$^{\circ}$}\rlap{\hskip.11\wd0 .}$^{\circ}
          \else \setbox0=\hbox{$^{\circ}$}\rlap{\hskip.11\wd0 .}$^{\circ}$\fi}
\def\arcs{\ifmmode {^{\scriptstyle\prime\prime}}
          \else $^{\scriptstyle\prime\prime}$\fi}
\def\arcm{\ifmmode {^{\scriptstyle\prime}}
          \else $^{\scriptstyle\prime}$\fi}
\newdimen\sa  \newdimen\sb
\def\parcs{\sa=.07em \sb=.03em
     \ifmmode \hbox{\rlap{.}}^{\scriptstyle\prime\kern -\sb\prime}\hbox{\kern -\sa}
     \else \rlap{.}$^{\scriptstyle\prime\kern -\sb\prime}$\kern -\sa\fi}
\def\parcm{\sa=.08em \sb=.03em
     \ifmmode \hbox{\rlap{.}\kern\sa}^{\scriptstyle\prime}\hbox{\kern-\sb}
     \else \rlap{.}\kern\sa$^{\scriptstyle\prime}$\kern-\sb\fi}
\def\ra[#1 #2 #3.#4]{#1\sup{h}#2\sup{m}#3\sup{s}\llap.#4}
\def\dec[#1 #2 #3.#4]{#1\deg#2\arcm#3\arcs\llap.#4}
\def\deco[#1 #2 #3]{#1\deg#2\arcm#3\arcs}
\def\rra[#1 #2]{#1\sup{h}#2\sup{m}}

\def\dots{\relax\ifmmode \ldots\else $\ldots$\fi}
\def\WHzsr{\ifmmode $W\,Hz\mo\,sr\mo$\else W\,Hz\mo\,sr\mo\fi}
\def\mHz{\ifmmode $\,mHz$\else \,mHz\fi}
\def\GHz{\ifmmode $\,GHz$\else \,GHz\fi}
\def\mKs{\ifmmode $\,mK\,s$^{1/2}\else \,mK\,s$^{1/2}$\fi}
\def\muKs{\ifmmode \,\mu$K\,s$^{1/2}\else \,$\mu$K\,s$^{1/2}$\fi}
\def\muKRJs{\ifmmode \,\mu$K$_{\rm RJ}$\,s$^{1/2}\else \,$\mu$K$_{\rm RJ}$\,s$^{1/2}$\fi}
\def\muKHz{\ifmmode \,\mu$K\,Hz$^{-1/2}\else \,$\mu$K\,Hz$^{-1/2}$\fi}
\def\MJysr{\ifmmode \,$MJy\,sr\mo$\else \,MJy\,sr\mo\fi}
\def\MJysrmK{\ifmmode \,$MJy\,sr\mo$\,mK$_{\rm CMB}\mo\else \,MJy\,sr\mo\,mK$_{\rm CMB}\mo$\fi}
\def\microns{\ifmmode \,\mu$m$\else \,$\mu$m\fi}

\def\muK{\ifmmode \,\mu$K$\else \,$\mu$\hbox{K}\fi}
\def\microK{\ifmmode \,\mu$K$\else \,$\mu$\hbox{K}\fi}
\def\muW{\ifmmode \,\mu$W$\else \,$\mu$\hbox{W}\fi}
\def\kms{\ifmmode $\,km\,s$^{-1}\else \,km\,s$^{-1}$\fi}
\def\kmsMpc{\ifmmode $\,\kms\,Mpc\mo$\else \,\kms\,Mpc\mo\fi}
\providecommand{\sorthelp}[1]{}

\def\WMAP{\textit{WMAP}}

\renewcommand{\L}[0]{\tens{L}}

\newcommand{\BP}{\textsc{BeyondPlanck}}

\def\inv{^{-1}}

\begin{document}

\title{\textsc{BeyondPlanck} V. Minimal ADC Corrections for \Planck\ LFI}
\newcommand{\oslo}[0]{1}
\newcommand{\manch}[0]{2}
\newcommand{\milanoA}[0]{3}
\newcommand{\milanoB}[0]{4}
\newcommand{\milanoC}[0]{5}
\newcommand{\triesteB}[0]{6}
\newcommand{\planetek}[0]{7}
\newcommand{\princeton}[0]{8}
\newcommand{\jpl}[0]{9}
\newcommand{\helsinkiA}[0]{10}
\newcommand{\helsinkiB}[0]{11}
\newcommand{\nersc}[0]{12}
\newcommand{\haverford}[0]{13}
\newcommand{\mpa}[0]{14}
\newcommand{\triesteA}[0]{15}
\author{\small
D.~Herman\inst{\oslo}\thanks{Corresponding author: D.~Herman; \url{daniel.herman@astro.uio.no}}
\and
R.~A.~Watson\inst{\manch}
\and
K.~J.~Andersen\inst{\oslo}
\and
\textcolor{black}{R.~Aurlien}\inst{\oslo}
\and
\textcolor{black}{R.~Banerji}\inst{\oslo}
\and
M.~Bersanelli\inst{\milanoA, \milanoB, \milanoC}
\and
S.~Bertocco\inst{\triesteB}
\and
M.~Brilenkov\inst{\oslo}
\and
M.~Carbone\inst{\planetek}
\and
L.~P.~L.~Colombo\inst{\milanoA}
\and
H.~K.~Eriksen\inst{\oslo}
\and
\textcolor{black}{M.~K.~Foss}\inst{\oslo}
\and
C.~Franceschet\inst{\milanoA,\milanoC}
\and
\textcolor{black}{U.~Fuskeland}\inst{\oslo}
\and
S.~Galeotta\inst{\triesteB}
\and
M.~Galloway\inst{\oslo}
\and
S.~Gerakakis\inst{\planetek}
\and
E.~Gjerl{\o}w\inst{\oslo}
\and
\textcolor{black}{B.~Hensley}\inst{\princeton}
\and
M.~Iacobellis\inst{\planetek}
\and
M.~Ieronymaki\inst{\planetek}
\and
\textcolor{black}{H.~T.~Ihle}\inst{\oslo}
\and
J.~B.~Jewell\inst{\jpl}
\and
\textcolor{black}{A.~Karakci}\inst{\oslo}
\and
E.~Keih\"{a}nen\inst{\helsinkiA, \helsinkiB}
\and
R.~Keskitalo\inst{\nersc}
\and
G.~Maggio\inst{\triesteB}
\and
D.~Maino\inst{\milanoA, \milanoB, \milanoC}
\and
M.~Maris\inst{\triesteB}
\and
A.~Mennella\inst{\milanoA, \milanoB, \milanoC}
\and
S.~Paradiso\inst{\milanoA, \milanoB}
\and
B.~Partridge\inst{\haverford}
\and
M.~Reinecke\inst{\mpa}
\and
A.-S.~Suur-Uski\inst{\helsinkiA, \helsinkiB}
\and
T.~L.~Svalheim\inst{\oslo}
\and
D.~Tavagnacco\inst{\triesteB, \triesteA}
\and
H.~Thommesen\inst{\oslo}
\and
D.~J.~Watts\inst{\oslo}
\and
I.~K.~Wehus\inst{\oslo}
\and
A.~Zacchei\inst{\triesteB}
}
\institute{\small
Institute of Theoretical Astrophysics, University of Oslo, Blindern, Oslo, Norway\goodbreak
\and
Jodrell Bank Centre for Astrophysics, Department of Physics and Astronomy, The University of Manchester, Manchester M13 9PL, UK
\and
Dipartimento di Fisica, Universit\`{a} degli Studi di Milano, Via Celoria, 16, Milano, Italy\goodbreak
\and
INAF-IASF Milano, Via E. Bassini 15, Milano, Italy\goodbreak
\and
INFN, Sezione di Milano, Via Celoria 16, Milano, Italy\goodbreak
\and
INAF - Osservatorio Astronomico di Trieste, Via G.B. Tiepolo 11, Trieste, Italy\goodbreak
\and
Planetek Hellas, Leoforos Kifisias 44, Marousi 151 25, Greece\goodbreak
\and
Department of Astrophysical Sciences, Princeton University, Princeton, NJ 08544,
U.S.A.\goodbreak
\and
Jet Propulsion Laboratory, California Institute of Technology, 4800 Oak Grove Drive, Pasadena, California, U.S.A.\goodbreak
\and
Department of Physics, Gustaf H\"{a}llstr\"{o}min katu 2, University of Helsinki, Helsinki, Finland\goodbreak
\and
Helsinki Institute of Physics, Gustaf H\"{a}llstr\"{o}min katu 2, University of Helsinki, Helsinki, Finland\goodbreak
\and
Computational Cosmology Center, Lawrence Berkeley National Laboratory, Berkeley, California, U.S.A.\goodbreak
\and
Haverford College Astronomy Department, 370 Lancaster Avenue,
Haverford, Pennsylvania, U.S.A.\goodbreak
\and
Max-Planck-Institut f\"{u}r Astrophysik, Karl-Schwarzschild-Str. 1, 85741 Garching, Germany\goodbreak
\and
Dipartimento di Fisica, Universit\`{a} degli Studi di Trieste, via A. Valerio 2, Trieste, Italy\goodbreak
}

\authorrunning{Herman et al.}
\titlerunning{BeyondPlanck LFI ADC Corrections}

\abstract{We describe the correction procedure for Analog-to-Digital
  Converter (ADC) differential nonlinearities (DNL) adopted in the
  Bayesian end-to-end \BP\ analysis framework. This method is nearly
  identical to that developed for the official LFI Data Processing
  Center (DPC) analysis, and relies on the binned rms noise profile of
  each detector data stream. However, rather than building the correction
  profile directly from the raw rms profile, we first fit a Gaussian
  to each significant ADC-induced rms decrement, and then derive the
  corresponding correction model from this smooth model. The main
  advantage of this approach is that only samples which are
  significantly affected by ADC DNLs are corrected. The
  new corrections are only applied to data for which there is a
  clear detection of the nonlinearities, and for which they perform
  at least comparably with the DPC corrections. Out of a total of 88 
  LFI data streams (sky and reference load for each of the 44 detectors) 
  we apply the new minimal ADC corrections in 25 cases, and maintain the
  DPC corrections in 8 cases. All these corrections are applied to 44 or
  70 GHz channels, while, as in previous analyses, none of the 30 GHz ADCs
  show significant evidence of nonlinearity. By comparing the \BP\ and DPC ADC
  correction methods, we estimate that the residual ADC uncertainty is
  about two orders of magnitude below the total noise of both the 44
  and 70\,GHz channels, and their impact on current cosmological
  parameter estimation is small. However, we also show that
  non-idealities in the ADC corrections can generate sharp stripes in
  the final frequency maps, and these could be important for future
  joint analyses with HFI, \WMAP, or other datasets. We therefore
  conclude that, although the existing corrections are adequate for
  LFI-based cosmological parameter analysis, further
  work on LFI ADC corrections is still warranted. }


\maketitle

\section{Introduction}
\label{sec:introduction}

The goal of the \BP\ project \citep{bp01} is to develop a computational framework for end-to-end Bayesian analysis for microwave and sub-mm experiments that accounts for both astrophysical and instrumental parameters, and apply this to the \textit{Planck} Low Frequency Instrument (LFI) data \citep{planck2016-l01,planck2016-l02}. The LFI consists of an array of radiometers which measure intensity and linear polarization in three frequency bands centered at 30, 44, and 70\,GHz. In each radiometer, the input signals from the sky and from a stable reference load at 4.5\,K are mixed, amplified, and detected by two detector diodes. Each diode receives alternately the signal from the sky and from the reference load, with a modulation rate of 4096\,Hz provided by a phase switch. Therefore, the output of each diode is a sequence of ($V_{\rm sky},\,V_{\rm ref}$) samples. The analog signal from each diode is then transferred in the Data Acquisition Electronics (DAE), where it is digitized by a 14-bit Analog-to-Digital Converter (from here out shortened to ADC). The scientific information is contained in the calibrated difference between sky and reference load signals. 

\begin{figure}[t]
\includegraphics[width=\linewidth]{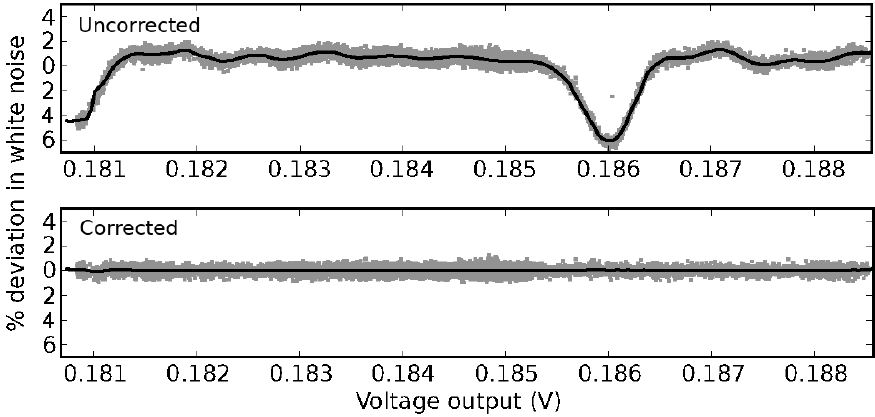}
\caption{Comparison of the uncorrected (\emph{top panel}) and
  corrected (\emph{bottom panel}) rms profile for the data stream 25M-sky01 (the sky voltage for detector 25M-01) as evaluated by the LFI DPC analysis; this figure is a reproduction
  of Figs.~10 and 11 in \citet{planck2013-p02a}. The main difference
  between the DPC procedure and the new approach is that the voltage
  ranges between the Gaussian dips are left unmodified by the updated
  algorithm.  }
\label{fig:DPC_correction}
\end{figure}

Ideally, the signal recorded by the ADC should depend linearly on the input voltage. However, significant ADC nonlinearities were noted early on in the \Planck\ data analysis, both in the LFI \citep{planck2013-p02,planck2013-p02a} and in the HFI \citep{planck2013-p03,planck2014-a08} instruments. In the case of LFI, the effect was first observed in the form of time-dependent deviation in the gain that lasted for several days, occurring at different times for different detectors within the same radiometer. Closer analysis revealed that these deviations were correlated with the white-noise level of the total-power sky/reference signal, as shown in the top panel Fig.~\ref{fig:DPC_correction}; this figure is a direct reproduction of Fig.~10 in \citet{planck2013-p02a}. The \Planck\ LFI team concluded that the most probable explanation was a slight increase in the quantization level of the signal in the form of  differential nonlinearity (DNL) in the ADCs \citep{planck2013-p02a,planck2014-a04}. The origin and impact of DNLs are well understood and documented by ADC chip producers.\footnote{See, e.g., \url{https://www.ti.com/lit/an/slaa013/slaa013.pdf} for a typical example.}

It should be noted that DNLs introduce a subtle effect that is independent of gain errors. The pseudo-correlation architecture of the LFI radiometers \citep{bersanelli2010} is such that amplifier gain fluctuations affect both sky and reference voltages in the same way, so that their effect is largely removed when taking the difference. However, ADC nonlinearities affect the sky and reference load signals independently. This is because the amplitudes of $V_{\rm sky}$ and $V_{\rm ref}$ are typically different and they will therefore reach the ADC electronics in two somewhat different ranges. Therefore, if not directly corrected for, the DNL will propagate directly into the differenced timestreams.

The \Planck\ LFI Data Processing Center (DPC) implemented two types of corrections for these nonlinearities. The first of these adopted the binned rms profile in Fig.~\ref{fig:DPC_correction} as a direct tracer of the effect, and defined a non-linear correction based on this profile that effectively flattened the rms profile. The rms profile result of this correction is shown in the bottom panel of Fig.~\ref{fig:DPC_correction}; see \citet{planck2013-p02a} for details. The second method addressed the problem from a more fundamental point-of-view, and defined a correction per bit. However, this approach was only shown to offer substantial improvement at 30\,GHz, but poor fits at 44 and 70\,GHz \citep{planck2016-l02}.

The starting point of the current paper is the first of the two methods, which is used for the official LFI DPC processing. In particular, we note in the following that the DPC corrections affect \emph{all} samples, as illustrated in Fig.~\ref{fig:DPC_correction}, and not only those associated with the rms decrements: The post-corrected rms profile is completely flat for all voltages. As a result, it is reasonable to assume that the DPC correction procedure may filter out effects that are unrelated to ADC nonlinearity, for instance correlated noise, thermal drifts, or actual sky signal. In this paper, we therefore introduce a small variation to the official procedure, in that we first fit a Gaussian to each decrement, and then use the resulting smooth function to define the ADC correction. The result is a correction that minimizes the impact with respect to non-ADC-related effects. 

To facilitate the following discussion we define here the naming convention, which follows the one used by the \Planck\ Collaboration. The LFI array comprises of 11 feedhorns\footnote{6 at 70\,GHz, 3 and 44\,GHz and 2 at 30\,GHz.}, which for historical reasons are numbered 18 to 28. Each horn feeds two radiometers, denoted "M" and "S"\footnote{The paired radiometers are connected to the "main" (M) or "side" (S) arm of an orthomode transducer, and therefore are orthogonal linear polarizations.}. Each radiometer has two diodes, denoted with 00--01 (M radiometer) and 10--11 (S radiometer). Each diode receives both the sky and reference load signals, alternating at about 4\,kHz, and thus produces two "data streams". The convention is that, for example, 25M-sky01 indicates the sky voltage data stream from horn 25, radiometer M, diode 01. In LFI there are a total of 88 data streams (44 diodes, sky and reference load) that need to be examined for ADC corrections.

The rest of the paper is organized as follows: Section~\ref{sec:framework} reviews the mathematical framework of the official DPC correction method, and introduces the practical details of the new implementation within the \BP\ framework. Section~\ref{sec:results} summarizes the effect of the ADC corrections, both at the level of binned rms profiles and frequency sky maps. Concluding remarks on the implementation and results are given in Sect.~\ref{sec:conclusions}.

\section{Methodology}\label{sec:framework}

In this section we first review the mathematical framework introduced by the \Planck\ DPC to correct for ADC DNLs \citep{planck2013-p02}, and then discuss the modifications adopted in the \BP\ framework. We note that the material presented in Sect.~\ref{sub:math} was derived in its entirety within the \Planck\ LFI DPC, but it was never detailed in the \Planck\ papers, only in internal documents. The current presentation therefore represents the first complete and publicly accessible summary of this method.

\subsection{DPC correction procedure}\label{sub:math}

For an ideal and perfectly linear ADC, the voltage step width corresponding to an increase in the analog-to-digital units (ADUs) should be 1 Least Significant Bit (LSB). An ADC DNL occurs when the spacing between some output ADUs do not correspond exactly to 1 LSB, altering the quantization step of the incoming signal. Considering this in terms of the input voltage $V$ that the instrument digitizes through the ADC, as quantified by some response function $R(V)$, the nonlinearity causes a decreased response for the offending bit.

\begin{figure}[t]
\includegraphics[width=\linewidth]{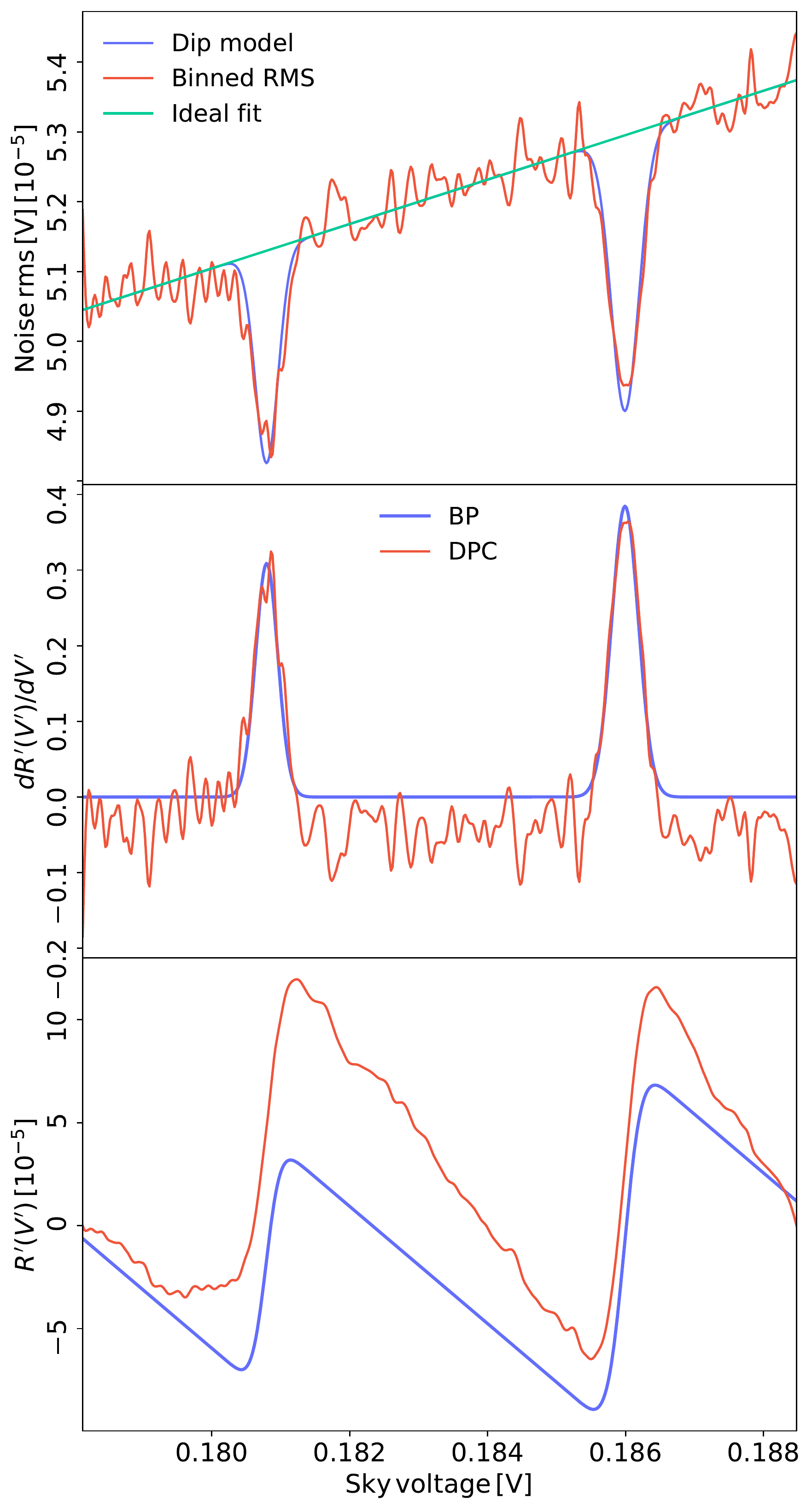}
\caption{(\emph{Top}:) Raw binned rms profile, $\delta V'$ (red), compared with a Gaussian fit (blue) and an ideal linear function (green) for a single data stream. (\emph{Middle}:) Inverse differential response function, $dR'(V')/dV'$, for the same data stream, as computed with the DPC (red) and \BP\ (blue) approaches.  (\emph{Bottom}:) Constructed inverse response function, $R'(V')$, for both DPC and \BP.}
\label{fig:adc_pipeline}
\end{figure}

The LFI ADCs convert measured voltages into 14-bit integers. In an ideal world, the conversion from the true voltage $V$ to the apparent (recorded) voltage $V'$ should follow a perfectly linear response $R'(V')$ given by
\begin{equation}
V = V'R'(V') = G(t)\,T_{\rm sys},
\label{eq:perfect}
\end{equation}
where $T_{\rm sys}$ is the system temperature and $G(t)$ is the radiometer gain. A small variation in the system temperature $\delta T$, caused either by a sky signal change or a noise fluctuation, leads to an increased detector voltage $\delta V$ and an altered apparent voltage $\delta V'$,
\begin{equation}
V+\delta V = (V' + \delta V')\,R'(V' + \delta V') = G(t)\,(T_{\rm sys}+\delta T).
\label{eq:perturbation}
\end{equation}
Assuming that the gain and the system temperature are fixed over the small signal measurement duration, we differentiate with respect to $V'$ and obtain
\begin{equation}
\delta V = \Big(V' \frac{dR'(V')}{dV'} + R'(V')\Big) \, \delta V' = G(t)\, \delta T.
\label{eq:diff}
\end{equation}
In addition to the main response $R'(V')\delta V'$, there is an additional $V' \frac{dR'(V')}{dV'}$ term, which is dependent on the differential of the response curve $R'(V')$. Since only $\delta V'$ and $V'$ are directly observable, we express the differential in terms of these variables,
\begin{equation}
V' = \frac{G(t)T_{\rm sys}}{R'(V')},
\label{eq:vprime}
\end{equation}
and
\begin{equation}
\delta V' = \frac{G(t)\delta T}{V' \frac{dR'(V')}{dV'} + R'(V')}.
\label{eq:delvprime}
\end{equation}
Substituting Eq.~\eqref{eq:vprime} into Eq.~\eqref{eq:delvprime} via the gain factor $G(t)$,  we find
\begin{align}
\delta V' &= \frac{R'(V')\,V'\,\delta T}{T_{\rm sys}}\Big[V' \frac{dR'(V')}{dV'} + R'(V')\Big]^{-1}\\
&= \frac{R'(V')V'}{V' \frac{dR'(V')}{dV'} + R'(V')}\frac{\delta T}{T_{\rm sys}}\\
&= \frac{a\,R'(V')V'}{V' \frac{dR'(V')}{dV'} + R'(V')},
\label{eq:delvprime2}
\end{align}
where $a\equiv\delta T/T_{\rm sys}$, which is an unknown constant corresponding to the predicted slope of the diode ADC. In the case of a linear response function (i.e. $dR'(V')/dV' = 0$), we find a direct proportionality between $\delta V'$ and $V'$, while the differential term in the denominator induces nonlinearities in the conversion between the two.

Departures from linearity can be quantified in terms of the change in the response function, $dR'(V')/dV'$, and this function is called the inverse differential response function (IDRF),
\begin{equation}
\frac{dR'(V')}{dV'} = \frac{a\,R'(V')}{\delta V'} - \frac{R'(V')}{V'} = \Big(\frac{a}{\delta V'} - \frac{1}{V'} \Big)\, R'(V').
\label{eq:idrf}
\end{equation}
By convention, we require the IDRF to have vanishing baseline and slope; these quantities are perfectly degenerate with the absolute instrument gain, and setting these to zero decouples the overall ADC correction from the absolute gain as far as possible. (We note that the ADC corrections still couple significantly to time-variable gain fluctuations, as discussed in Sect.~\ref{sec:results}.) 

Using our expression for $dR'(V')/dV'$, we get the final reconstructed inverse response function (RIRF) by the IDRF with respect to the voltage $V'$. At a voltage $V'_k$, 
\begin{equation}
R'(V'_k) = \frac{dR'(V'_k)}{dV'}\Delta V' + R'(V'_{k-1}),
\label{eq:rirf}
\end{equation}
where $\Delta V'$ is the bin width. For the first bin, we choose ${R'(V'_0)=1}$, once again to decorrelate the ADC corrections from the absolute gain. With this expression in hand, the final reconstructed voltage may be written as
\begin{equation}
  V = V'R'(V') + V'.
  \label{eq:tot_correction}
\end{equation}
In practice, $R'(V')$ is stored as a regular cubic spline, which ensures that the corrections may be applied both accurately and efficiently.

Since all of the above corrections are numerically small, it is is useful to define the following functions for visual inspection. The first auxiliary function is
\begin{equation}
\alpha = \frac{\delta V'}{aV'}-1.
\label{eq:alpha_def}
\end{equation}
$\alpha$ represents the departure of the binned rms from the expected rms of the data, and is conveniently plotted in units of percent. Second, we define
\begin{equation}
\beta = V/V'-1
\label{eq:beta_def}
\end{equation}
to be the relative deviation from linearity in the actual ADC correction, and this is typically of the order of $10^{-5}$.

\begin{figure}[t!]
\includegraphics[width=\linewidth]{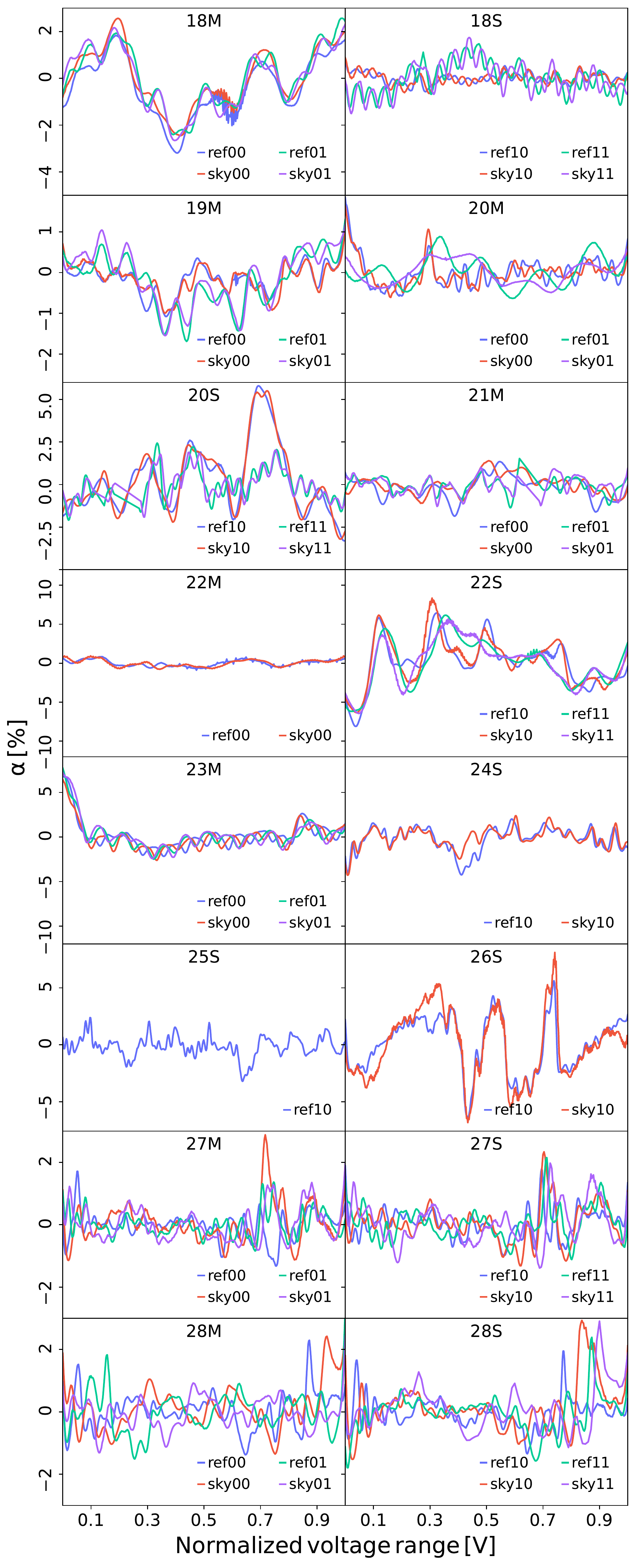}
\caption{Binned rms for all uncorrected data streams in the 30, 44, and 70\,GHz channels, separated into panels based on data stream names. The panels labeled 27M, 27S, 28M, and 28S correspond to the 30\,GHz radiometers.}
\label{fig:uncorrected}
\end{figure}

\begin{figure*}[p]
\includegraphics[width=\linewidth]{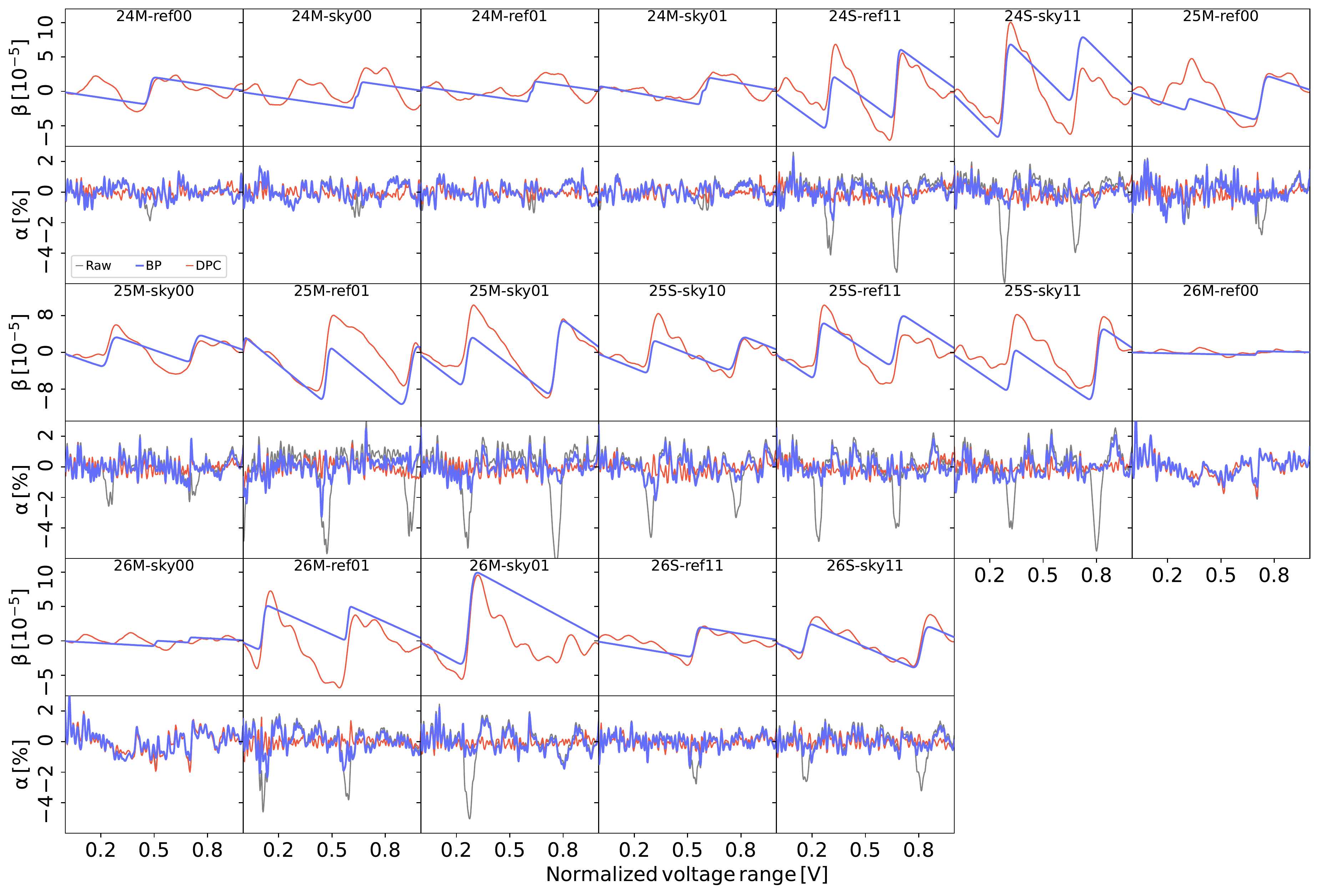}
\caption{ADC nonlinearity correction in the 44\,GHz channel is applied to 19 out of 24 data streams. (\emph{top panels}): Normalized ADC correction tables; (\emph{bottom panels}): Normalized binned rms. Red curves correspond to DPC results, while blue curves correspond to the minimal corrections derived in this work. The raw rms profiles are plotted in grey. All of the data streams which are corrected within the 44\,GHz channel adopt the minimal corrections derived here.}\label{fig:44ghz_corrected}
\vspace*{6mm}

\includegraphics[width=\linewidth]{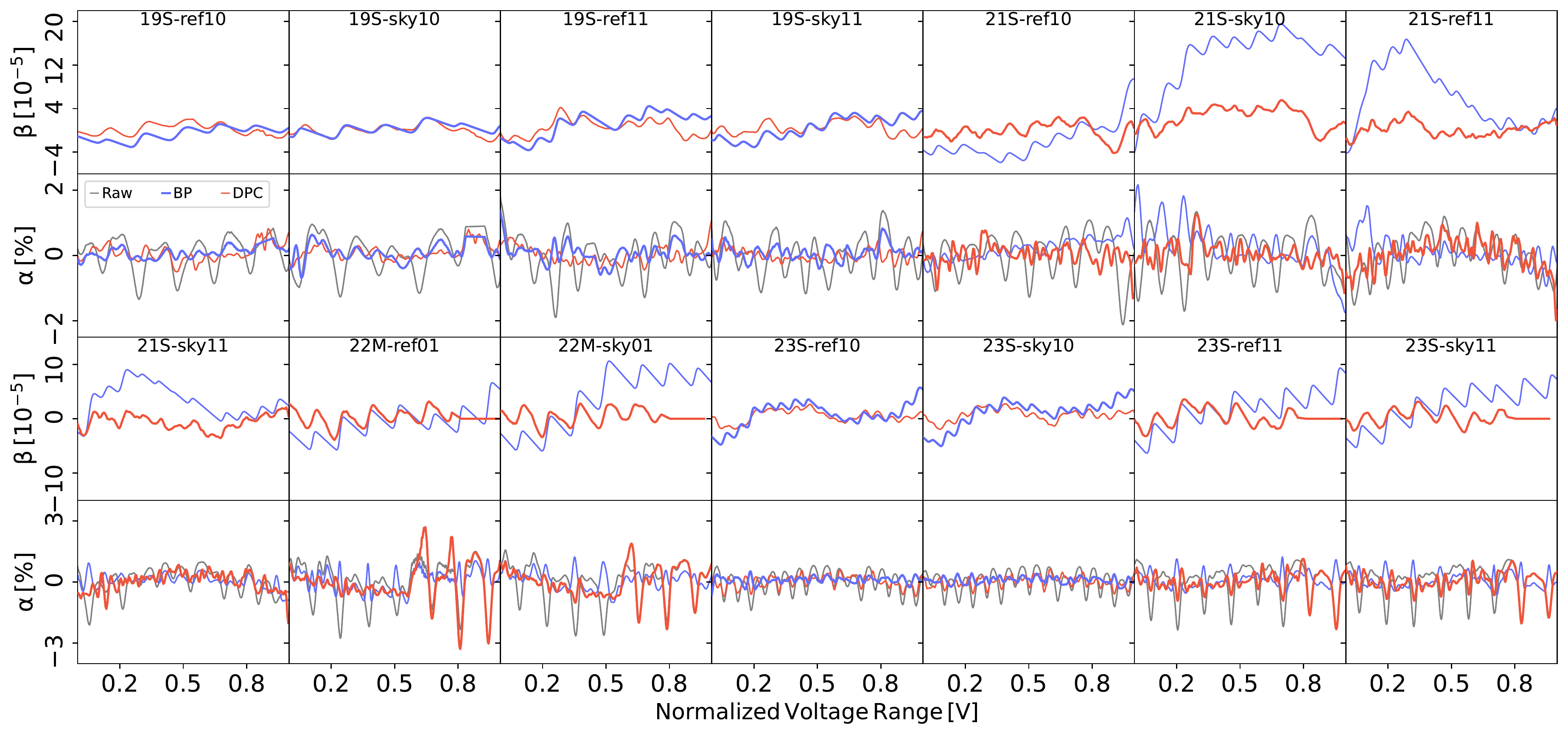}
\caption{Same as Fig.~\ref{fig:44ghz_corrected} for 70\,GHz (14 out of 48 data streams). The data streams corrected with the minimal \BP\ approach are shown with the thicker line in blue, while the DPC tables are adopted for the data streams with the red line as the thickest.}\label{fig:70ghz_corrected}
\end{figure*}

\subsection{Minimal ADC corrections through Gaussian fits}
\label{sub:implementation}

The red curves in Fig.~\ref{fig:adc_pipeline} show the primary quantities in the above derivation for 25M-sky01, which is the same data stream highlighted in Fig.~\ref{fig:DPC_correction} for the DPC analysis. In the top panel, the red curve shows the raw binned rms profile, $V'$,\footnote{The reasons this looks visually different than the top panel in Fig.~\ref{fig:DPC_correction} are different flagging, data selection and binning schemes.} while the red curve in the middle panel shows the IDRF. The red curve in the bottom panel shows the RIRF, which in practice defines the splined non-linear correction.

In these plots, there are substantial small-scale fluctuations that propagate from the raw binned rms profile into the final correction. These are not due to ADC nonlinearity, but rather caused by correlated noise, long-term system temperature drifts, and, potentially, sky signal, in particular from the CMB dipole. However, the DPC method outlined above does not discriminate between these various terms, but simply integrates over the full rms profile.

In this paper, we therefore introduce one additional step in the above procedure, and fit a 3-parameter Gaussian with free amplitude, location, and width to each significant Gaussian dip. This is illustrated as a blue curve in the top panel of Fig.~\ref{fig:adc_pipeline}. This smooth fitted curve is then used as input for the rest of the algorithm, eventually resulting in a smooth correction curve in the bottom panel. The advantage of this procedure is that only significant ADC nonlinearities are affected by the correction, leaving most of the non-affected samples unchanged (up to an overall linear re-scaling). As a result, we denote this procedure ``minimal'', indicating that we do not modify the data more than strictly necessary.

\begin{figure*}

\includegraphics[width=0.33\linewidth]{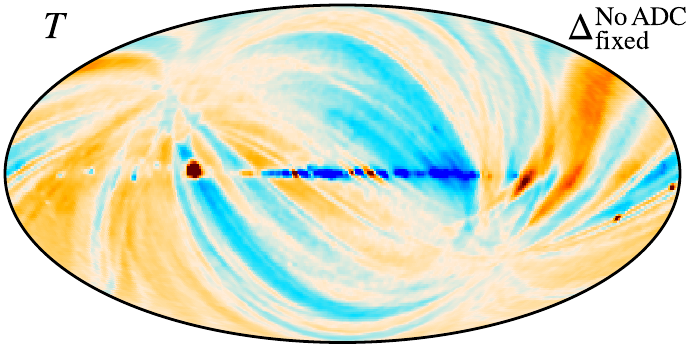}
\includegraphics[width=0.33\linewidth]{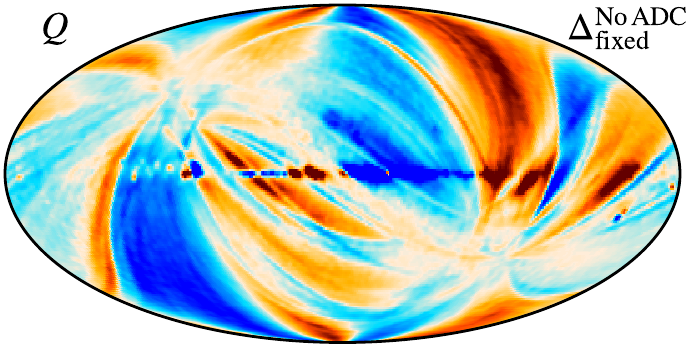}
\includegraphics[width=0.33\linewidth]{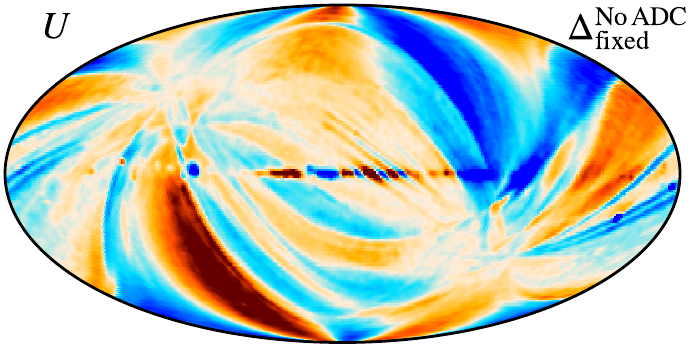}\\

\includegraphics[width=0.33\linewidth]{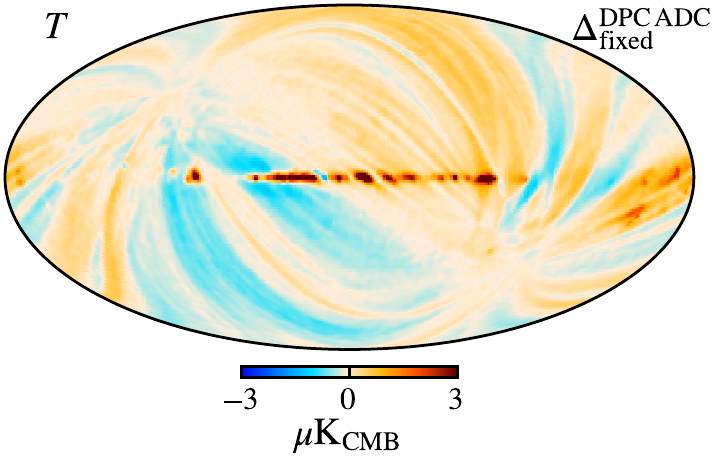}
\includegraphics[width=0.33\linewidth]{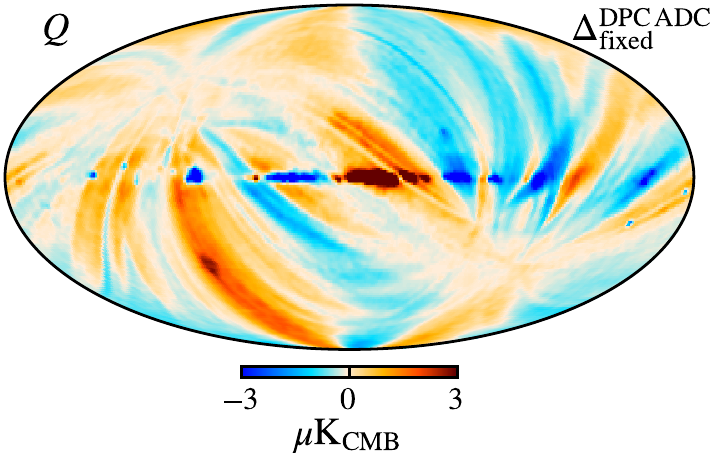}
\includegraphics[width=0.33\linewidth]{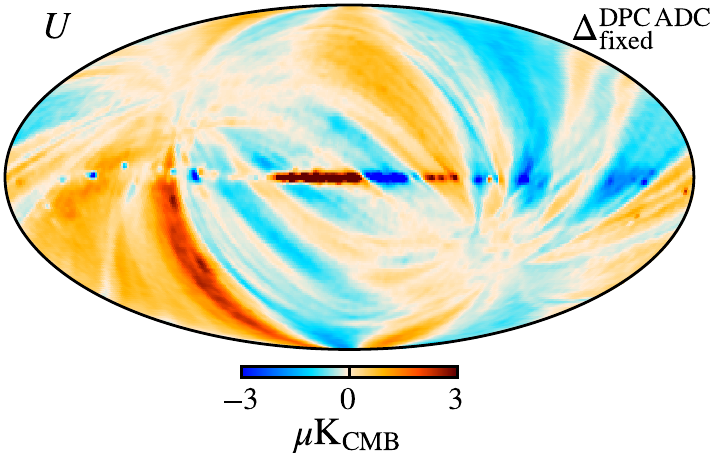}\\

\includegraphics[width=0.33\linewidth]{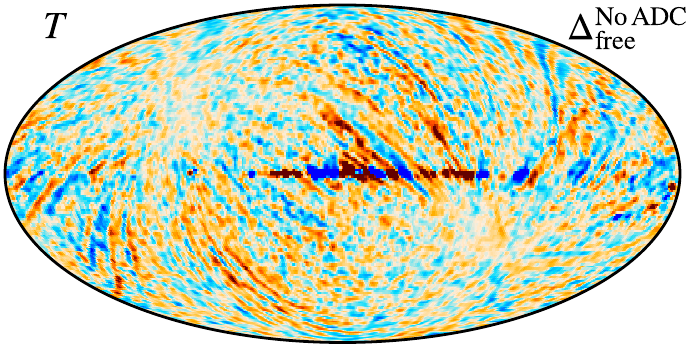}
\includegraphics[width=0.33\linewidth]{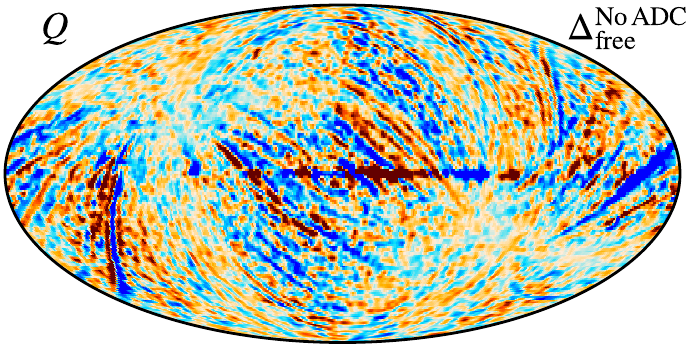}
\includegraphics[width=0.33\linewidth]{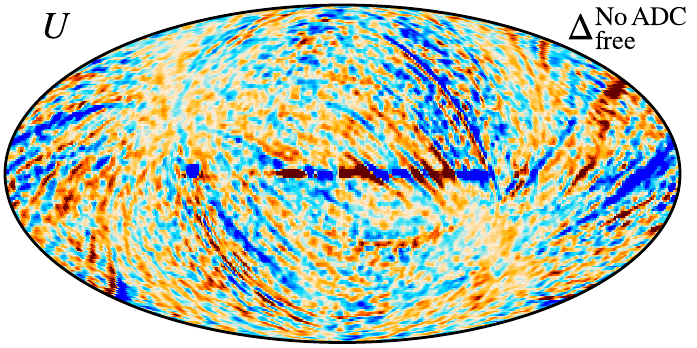}\\

\includegraphics[width=0.33\linewidth]{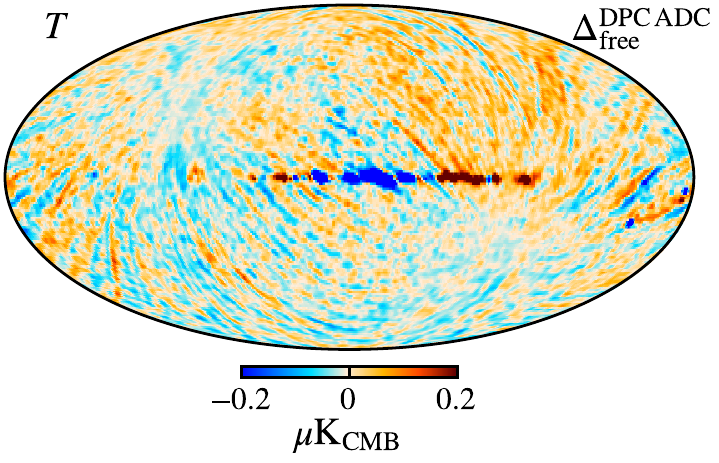}
\includegraphics[width=0.33\linewidth]{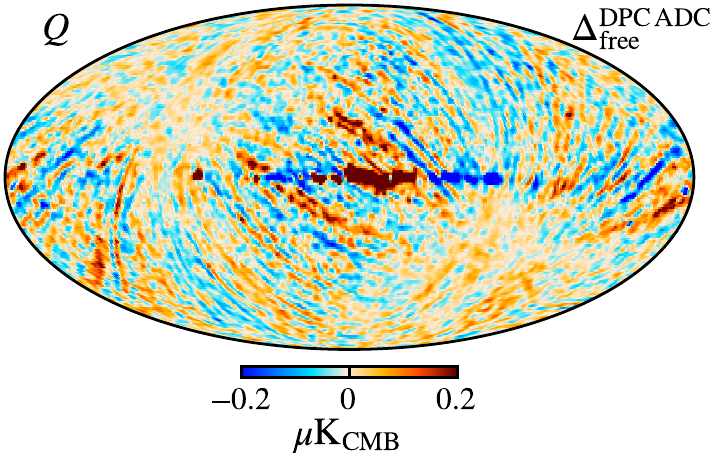}
\includegraphics[width=0.33\linewidth]{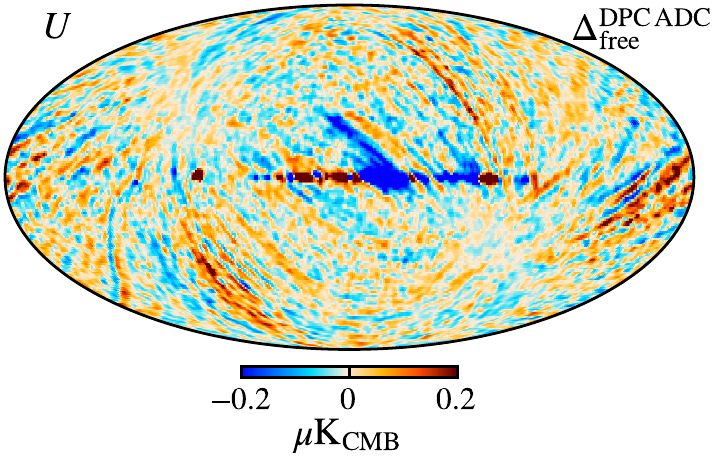}

\caption{Effect of the ADC corrections on the 44\,GHz sky maps. (\textit{Top row:}) Difference between the \BP\ ADC corrections and no ADC corrections with the gain solution fixed. (\textit{Top middle row:}) Difference between the \BP\ ADC corrections and no ADC corrections with gain sampling enabled. (\textit{Bottom middle row:}) Difference between the \BP\ ADC corrections and the DPC ADC corrections, with the gain solution fixed. (\textit{Bottom row:})  Difference between the \BP\ ADC corrections and the DPC ADC corrections, with gain sampling enabled. }
\label{fig:sky_maps_44}
\end{figure*}

Practically speaking, the first step in the process of implementing the minimal ADC corrections is to determine the absolute minimum and maximum voltage for each data stream. The full voltage range is then divided into 500 bins. As will be seen in Sect.~\ref{sub:binnedrms}, our voltage range and binning differ from the DPC correction tables, and this is primarily due to the different data selection and masking choices adopted in the DPC and \BP\ analyses \citep{bp01,bp06,bp10}. 

With the binning scheme in place, the next step is to estimate the binned rms profile. We do this for each data stream sample by estimating the local rms over a window centered on the current sample; we adopt a window size of 10 samples in the current analysis, but note that the results are not very sensitive to this choice. Voltage bins with fewer than 20 samples are masked out.

The third step is to identify significant rms deviations. We do this by searching through windows of 10 voltage bins for $2\,\sigma$ deviations. If a given bin is lower than all of its neighbors it is flagged as a dip. (In addition to this automatic flagging procedure, we have also inspected all candidates by eye to ensure that there are no spurious detections). Not all data streams return any significant dips, as defined by this procedure, and the rms profiles for those data streams are shown in Fig.~\ref{fig:uncorrected}. In particular, we note that no 30\,GHz data streams exhibit significant dips, while for 44 and 70\,GHz there are 7 and 32 data streams, respectively, that show no convincing evidence of ADC nonlinearity. We note that the official DPC analysis applied no corrections to any of these, based on the same observations.

The fourth step is to fit a 3-parameter Gaussian to each significant dip, with a free amplitude, location and width. This is done using a standard $\chi^2$ minimization, in which the standard deviation is defined by the variance between neighboring bins in the rms profile. With this smooth function in hand, the remaining procedure is identical to the DPC approach, and the inverse differential response function is integrated over the voltage range according to Eq.~\eqref{eq:rirf}.

As described by \citet{bp01}, the \BP\ project adopts a Gibbs sampling approach to LFI processing, and as a result this procedure should in principle be performed at each Gibbs step to take uncertainties in the dip fitting procedure into account. However, since the above fitting procedure operates with raw TOD, there is no feedback from higher-level stages in the analysis, for instance component separation \citep{bp13}. In this work we instead perform the corrections of the DNL as a strict preprocessing step, to save both memory and processing time. 

We also note that ADC uncertainties, while independent in origin, are highly degenerate with the time-dependent instrument gain, and ADC error propagation is therefore to a large extent already taken into account through the gain sampling procedure described by \citet{bp07}. 

Future work on the physical bit-based ADC correction procedure discussed by \citet{planck2016-l02} may rely on a full data model to constrain the ADC nonlinearity, and in this case coupling between the sky signal and ADC corrections would become more important.

\section{Results}\label{sec:results}

\begin{figure*}
\includegraphics[width=0.33\linewidth]{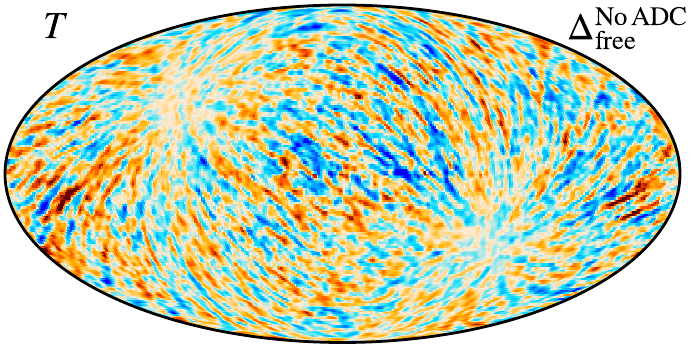}
\includegraphics[width=0.33\linewidth]{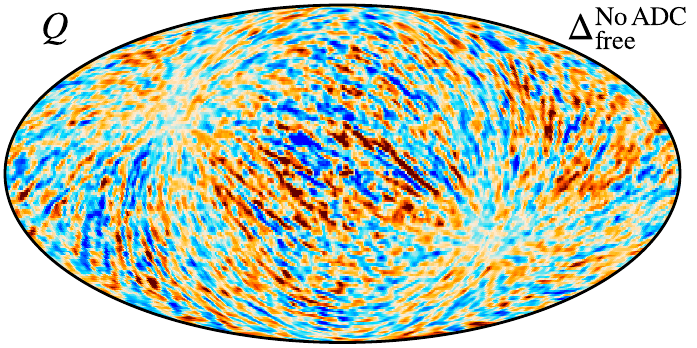}
\includegraphics[width=0.33\linewidth]{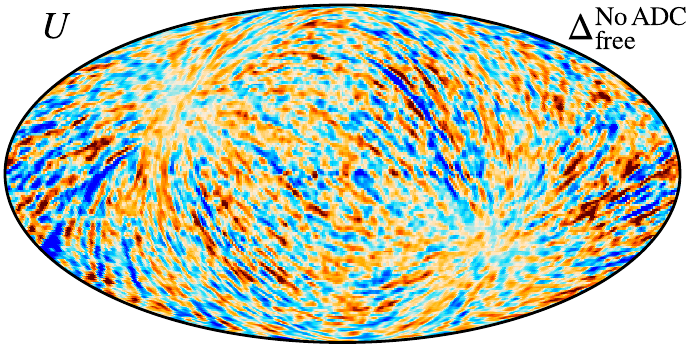}\\
\includegraphics[width=0.33\linewidth]{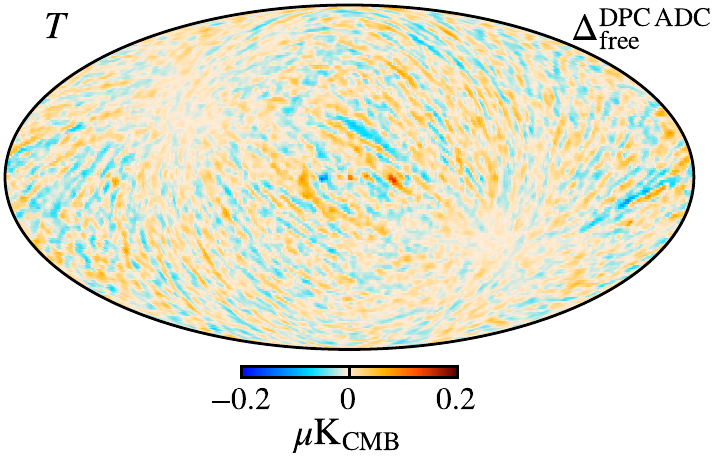}
\includegraphics[width=0.33\linewidth]{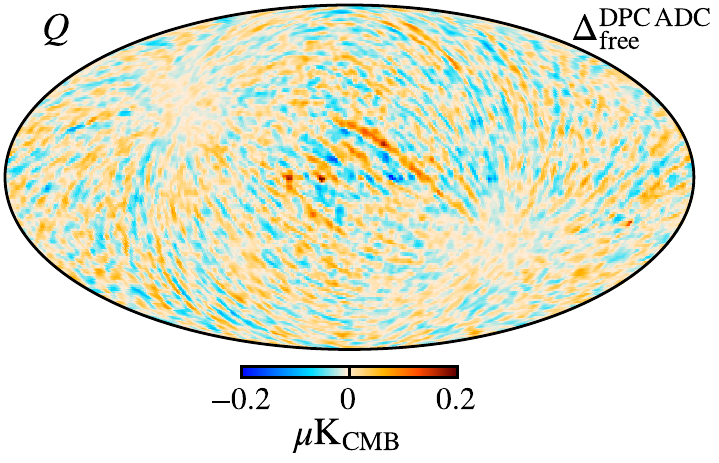}
\includegraphics[width=0.33\linewidth]{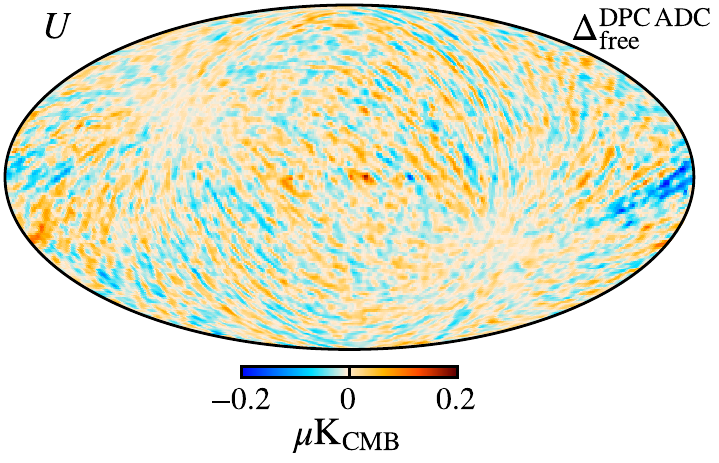}

\caption{Effect of ADC corrections on the 70\,GHz sky maps. (\textit{Top row:}) Difference between the \BP\ ADC corrections the no ADC corrections with gain sampling enabled. (\textit{Bottom row:}) Difference between the \BP\ ADC corrections and the DPC ADC corrections with gain sampling enabled.}

\label{fig:sky_maps_70}
\end{figure*}

\subsection{Binned rms profiles and ADC corrections}\label{sub:binnedrms}

We are now finally ready to present the LFI ADC nonlinearity corrections derived and used in this work. These are summarized in Figs.~\ref{fig:44ghz_corrected} and \ref{fig:70ghz_corrected} for all relevant 44 and 70\,GHz data streams, respectively; we recall that all the data streams listed in Fig.~\ref{fig:uncorrected} do not show any significant evidence of ADC nonlinearity, and they are therefore omitted in the following. In these figures, each data stream is summarized in terms of two panels that show the normalized correction, $\beta$ (top panel), and the normalized binned rms profile, $\alpha$ (bottom panel), respectively. In each case, the DPC results are shown as red curves, while the \BP\ results are shown as blue curves; in the bottom panel, the gray curve additionally shows the raw binned rms profile. 

Starting with the binned rms profiles, the first point to notice is the difference between the gray and colored lines: Clear dips in the gray curves indicate the presence of a significant ADC nonlinearity, while corresponding flat colored lines at the same locations indicate that the correction performs well. An example of this is 24S-ref11, which has two strong dips, and both of the correction procedures result in a binned profile that appears statistically consistent at the dip locations as measured with respect to the non-affected parts. In total, 19 out of 24 44\,GHz data streams show this type of behaviour. For the remaining 5 data streams, the ADCNL detection is marginal, and these could quite conceivably have been left uncorrected, similar to those shown in Fig.~\ref{fig:uncorrected} without major negative effects. In general, however, we conclude that the ADC corrections appear to perform well for the 44\,GHz channel.

The picture is more mixed regarding the 70\,GHz data streams. In many cases, we see that there are many dips spaced with regular intervals, for instance 23S-sky11; the latter observation strongly suggests that these are indeed due to ADC nonlinearities, rather than for instance thermal drifts or correlated noise. At the same time, we also see that some of the binned rms profiles do not appear linear, but rather parabolic (e.g., 21S-ref11) or with a discrete step (e.g., 22M-ref01). Whenever this happens, the \BP\ approach performs poorly, due to the strict assumption of localized nonlinearity imposed during the Gaussian fit. At the same time, it is also clear that the DPC approach is sub-optimal for many of these cases, as one can clearly see evidence of residual dips after correction (e.g., 22M-ref01).

For the current \BP\ processing, we adopt a conservative approach, and only apply the new minimal ADC corrections to data streams for which it is clear that they perform no worse than the DPC corrections. An arbitrary example of this is 26M-sky01. This data stream has one clear dip, and the minimal \BP\ correction appears fully statistically consistent with the rest of the profile at the dip location. At the same time, this correction clearly affects the data less than the DPC correction, as evidenced by the much larger overall non-ADC-induced scatter. An example of the contrary is 21S-sky10, for which the minimal correction shows clear evidence of over-corrected peaks at low voltages; in this case, it is unclear which correction is better. The correction procedure adopted for each data stream in the final \BP\ analysis is indicated by thick lines in Figs.~\ref{fig:44ghz_corrected} and \ref{fig:70ghz_corrected}. Here we see that we apply the minimal procedure to each of the 44\,GHz data streams that are corrected, while only 6 of the 14 corrected 70\,GHz data streams are corrected with the minimal procedure. For the other 8 corrected 70\,GHz data streams, the DPC corrections are adopted.

Looking at the normalized correction curves ($\beta$; top panels), the main difference between the two approaches is that the DPC corrections exhibit more small-scale variations than the minimal corrections. Eliminating these was precisely the main motivation of the minimal correction procedure introduced in this paper, reducing coupling with non-ADC-related effects.

\begin{figure*}
\includegraphics[width=0.5\linewidth]{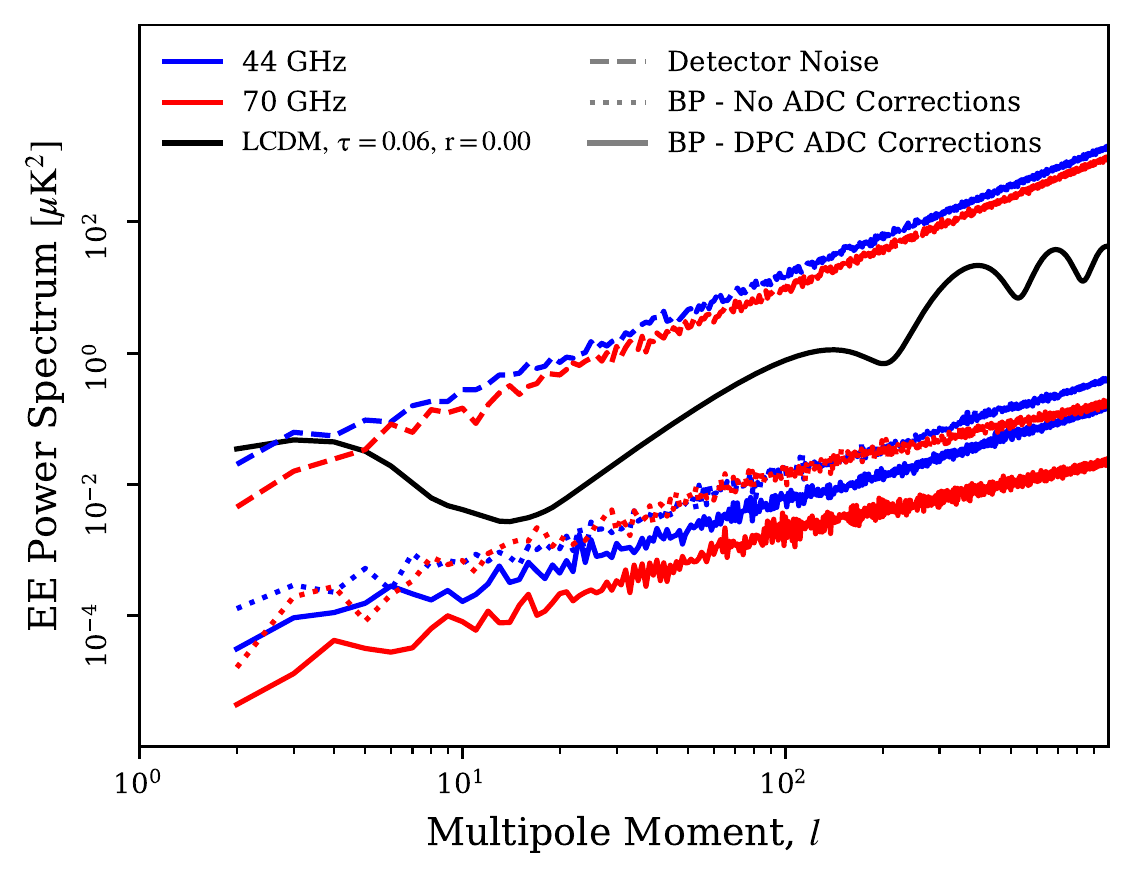}
\includegraphics[width=0.5\linewidth]{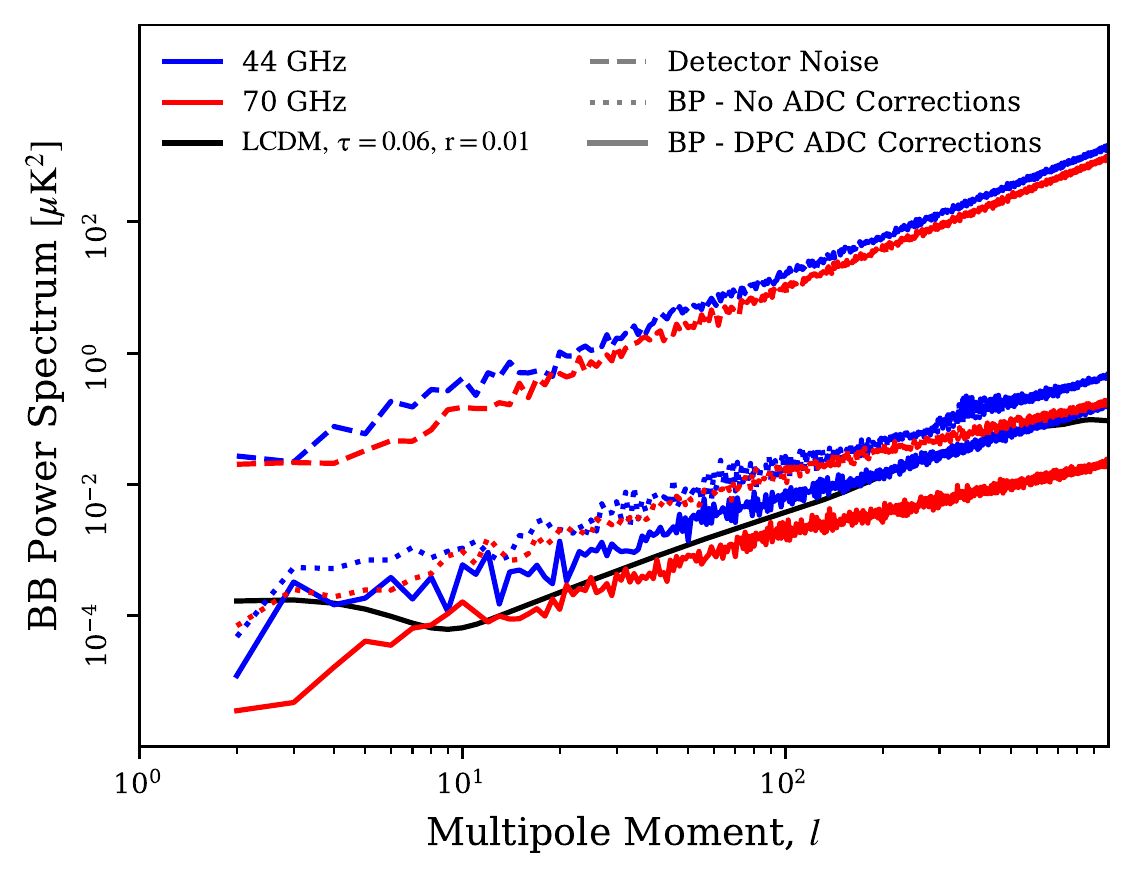}
\caption{Power spectrum comparisons for the EE (\textit{left}) and BB (\textit{right}) sky signals. The blue and red lines correspond to the 44 and 70 GHz detectors respectively. The power of the changes in the ADC corrections is significantly smaller than the noise power spectrum, yet of comparable power to the $\Lambda$CDM BB power spectrum.}\label{fig:ps}
\end{figure*}

\subsection{Sky map comparisons}\label{sub:sky_maps}

We conclude this section with a comparison of various ADC corrections and processings at the level of pixelized maps in order to bound their net impact on final frequency maps, as summarized in Fig.~\ref{fig:sky_maps_44} for the 44\,GHz channel. First, the top row shows the difference between \BP\ sky maps with and without ADC corrections applied, while keeping all other parameters in the data model fixed, including the gain. This difference map therefore provides an absolute upper limit on the potential impact of the ADC corrections at their most basic level. Generally speaking, we see that the differences are less than 10\muK, and with a morphology closely related to the scanning strategy. We also note that the effect is about twice as large in polarization as in temperature.

The second row shows a similar difference map between \BP\ 44\,GHz frequency maps when using either the DPC or \BP\ ADC corrections, but still fixing all other parameters at the same values. These differences are smaller by a factor of about two, but highly significant compared to the cosmological and astrophysical signals at the 44\,GHz channel. 

However, it is important to emphasize that in both of the above cases, all other instrumental parameters were fixed, and this applies in particular to the time-variable gain. As noted in Sect.~\ref{sub:math}, the ADC nonlinearity correction defined by Eq.~\eqref{eq:tot_correction} is strongly degenerate with the instrumental gain, $G(t)$, defined in Eq.~\eqref{eq:perfect} \citep{bp07}. Therefore, different ADC corrections may to a very large extent be accounted for by modifying the time-variable gain model. The third row in Fig.~\ref{fig:sky_maps_44} demonstrates this explicitly; in this case, we apply no ADC corrections at all, but do fit the gain. In this case, the resulting map agrees with the default \BP\ algorithm to about 0.2\muK\,over most of the sky. The irreducible residual effect of ADC corrections is seen as a few sharp stripes, and these coincide with the Pointing Period IDs (PIDs) for which the ADC correction, as defined by Eq.~\eqref{eq:rirf}, exhibits a large gradient; such changes cannot be accounted for with a single linear gain factor per PID.

Finally, the bottom panel shows a corresponding difference between the \BP\ and DPC algorithms after fitting gains in each case. These differences are lower than the previous by about a factor of two, similar to the fixed gain case, and most of the sharp stripes have vanished. At this point, the main residual effects include a small dipole difference in $T$ of about 0.1\muK, which is small compared to the full final CMB dipole uncertainty of 1\muK\, \citep{bp10}, a few sharp stripes following the scanning strategy, and an overall noise-like floor, which most likely is a second-order coupling between the gain and the correlated noise parameters \citep{bp06}.

Figure~\ref{fig:sky_maps_70} shows a similar comparison for the 70\,GHz channel, but for brevity only for the latter free gain cases. In this case, the main net effect of the ADC corrections is clearly a lower level of correlated noise.

\subsection{Power spectrum impact}\label{sub:ps}

We conclude this analysis with a brief discussion of the effect of ADC corrections on the polarization angular power spectra for each of the two frequency maps. These are shown in Fig.~\ref{fig:ps} for both $EE$ and $BB$ spectra, both computed over 95.4\,\% of the sky, after excluding the \BP\ processing mask \citep{bp06}. First, the dashed curves show the total correlated-plus-white noise power spectra for a single Gibbs sample, and we see that these cross the best-fit $\Lambda$CDM spectrum at $\ell\approx3$ and 5 for 44 and 70\,GHz, respectively, in agreement with previously reported LFI results \citep{planck2016-l02,bp11}.

In comparison, the dotted line shows the power spectrum of the difference maps obtained when applying no ADC corrections shown in Figs.~\ref{fig:sky_maps_44} and \ref{fig:sky_maps_70}. These are about one and a half orders of magnitude lower than the noise at low multipoles, and about two and a half orders of magnitude at high multipoles. In general, the spectra appear featureless, and therefore correspond to a colored noise contribution, in good agreement with the visual interpretation in Figs.~\ref{fig:sky_maps_44} and \ref{fig:sky_maps_70}.

Finally, the solid lines show the corresponding power spectra for the difference maps between the \BP\ and DPC ADC corrections, and these provide a useful estimate of the actual uncertainties in the two models. These are lower than both the $EE$ recombination peak and the total noise at $\ell=2$--8 by two orders of magnitude, and residual ADC uncertainties are therefore not likely to significantly affect the \BP\ estimates of the optical depth of reionization \citep{bp12}. At the same time, it is important to note that if LFI data (current or previous) are to be used in studies of extended reionization models that include $\ell=10$--20, then the residual ADC uncertainties are no longer negligible, and the power contribution may account for as much as 30\,\% of the cosmological signal.

The same considerations apply to potential studies of cosmological $BB$ power and the tensor-to-scalar ratio, as seen in the right panel of Fig.~\ref{fig:ps}. In this case, the residual ADC power at 44\,GHz is comparable to the expected cosmological signal for $r=0.01$ and lensing at all multipoles; for the 70\,GHz channel, it is lower by a factor of three or so. Of course, these levels are still about two orders of magnitude lower than the noise, and ADC residuals do therefore not play a significant role in the current \BP\ estimates of the tensor-to-scalar ratio, which find $r<0.9$ at 95\,\% confidence, as presented by \citet{bp12} using only $\ell=2$--8. However, if one were to derive a single $BB$ power estimate by co-adding over all multipoles up to, say, $\ell=1000$, the ADC contribution could become non-negligible. We therefore conclude that residual ADC uncertainties are negligible with respect to the current \BP\ cosmological parameter estimates, but note that external users should be aware of their existence, as they could potentially become relevant for more aggressive analyses that consider wider multipole ranges.

\section{Conclusions}
\label{sec:conclusions}

The main goal of this paper is to document the correction procedure for ADC nonlinearities employed in the \BP\ framework. This procedure follows closely the method developed by the LFI DPC for the official \Planck\ analysis, with one notable difference: Rather than deriving the correction function directly from raw binned data, we fit a low-dimensional parametric function prior to inversion, thereby greatly reducing the number of degrees of freedom in the model. As a result, potential spurious coupling between the true sky signal and the ADC corrections is reduced, and the probability of introducing spurious biases minimized. The current paper provides the most complete publicly accessible summary of the \Planck\ LFI DPC ADC correction procedure published to date, as much of this material has until now only been available in the form of unpublished internal work notes. This is likely to be useful for future re-analyses of the \Planck\ LFI measurements that aim to go even deeper than \BP, for instance in the form of a future joint \Planck--LiteBIRD analysis. 

While performing this work, we have confirmed many of the LFI DPC results. Most importantly, we note that the 44\,GHz channel exhibits the clearest signatures of ADC nonlinearity, and these are quite straightforward to correct for, while the 30\,GHz channel shows no evidence of ADC nonlinearity, and no corrections are therefore applied in this case. In contrast, and somewhat disconcertingly, many 70\,GHz data streams both show significant ADC signatures, and they are difficult to correct for. In the current analysis, the new procedure is applied to 25 (19 at 44\,GHz, and 6 at 70\,GHz) data streams, while the DPC procedure is applied to 8 data streams. The remaining 55 data streams (including the 30\,GHz data streams) are left uncorrected as they do not show strong evidence of ADC nonlinearities, yet the amount of structure observed in these data streams renders a firm conclusion difficult.

To estimate the net impact of these corrections on the final frequency maps, we compare the \BP\ corrections both with making no ADC corrections at all and with the official DPC corrections. Through these comparisons, we have shown that residual ADC uncertainties are small compared to the noise of both channels, typically about two orders of magnitude lower in terms of power spectra. In absolute terms, the residual ADC power accounts for about 30\,\% of the $EE$ $\Lambda$CDM power in the minimum between $\ell=10$--20, and it is comparable to the $BB$ power of a $\Lambda$CDM model with a tensor-to-scalar ratio of $r\approx0.01$. In addition to this diffuse noise-like power,  residual ADC artifacts can induce sharp stripes in the map, and if such localized features are found in the final frequency maps, then ADC corrections provide a plausible hypothesis for their origin.

In summation, we conclude that the current LFI ADC corrections, whether computed by \BP\ or the LFI DPC, are sufficient for current cosmological analysis. At the same time, it is clear that there is still further room for improvements, and the most promising approach for this may be to integrate the physically motivated bit-level model pioneered by \citet{planck2016-l02} into the \BP\ framework, and sample the free parameters as part of the joint data model. This should be explored in future work, ideally jointly with high signal-to-noise observations from HFI that may be used to further disentangle ADC, gain and correlated noise fluctuations.

\begin{acknowledgements}
  We thank Prof.\ Pedro Ferreira and Dr.\ Charles Lawrence for useful suggestions, comments and 
  discussions. We also thank the entire \Planck\ and \WMAP\ teams for
  invaluable support and discussions, and for their dedicated efforts
  through several decades without which this work would not be
  possible. The current work has received funding from the European
  Union’s Horizon 2020 research and innovation programme under grant
  agreement numbers 776282 (COMPET-4; \BP), 772253 (ERC;
  \textsc{bits2cosmology}), and 819478 (ERC; \textsc{Cosmoglobe}). In
  addition, the collaboration acknowledges support from ESA; ASI and
  INAF (Italy); NASA and DoE (USA); Tekes, Academy of Finland (grant
   no.\ 295113), CSC, and Magnus Ehrnrooth foundation (Finland); RCN
  (Norway; grant nos.\ 263011, 274990); and PRACE (EU).
\end{acknowledgements}

\bibliographystyle{aa}

\bibliography{Planck_bib,BP_bibliography}

\end{document}